\documentclass[final,5p,times,twocolumn,authoryear]{elsarticle}

\usepackage{amsmath,amssymb,amsfonts}
\usepackage{amsmath}
\usepackage{multirow}
\usepackage{url}
\providecommand{\abs}[1]{\lvert#1\rvert}
\newcommand\var[1]{\mathrm{Var}{(#1)}}

\newcommand\col[1]{{\color{black}#1}} 

\newcommand{\eye}{{\rm{i}}}
\usepackage{color}
\usepackage{lineno}

\journal{Coastal Engineering}

\begin{document}

\begin{frontmatter}
  
\title{ HF radar estimation of ocean wave parameters: second-order Doppler spectrum versus Bragg wave modulation approach}
\author[1]{Ver\'onica Morales-M\'arquez}
    \author[1]{Dylan Dumas}
\author[1]{Charles-Antoine Gu\'erin}
\affiliation[1]{organization={MIO (Univ Toulon, Aix-Marseille Univ, CNRS, IRD)},
  city={Toulon},
  country={France}}
\begin{abstract}
  We propose an original technique for the HF radar estimation of the main sea state parameters by exploiting the amplitude modulation of the radar signal time series. While the classical method for ocean wave measurement is based on the second-order ocean Doppler spectrum, this alternative approach uses the slow amplitude modulation of the Bragg wave in the radar signal. Using a nearby buoy and the WWIII model as ground truth, we apply this method to an annual set of HF radar data in the vicinity of Toulon (Mediterranean coast of France) and compare it with \col{the classical formulas proposed by Barrick (1977) to derive the main wave parameters from the Doppler spectrum.  Although the modulation approach is physically attractive and easier to implement, we find that the Doppler spectrum-based method is far superior at this stage in calculating the significant wave height and the peak wave frequency, even at long range.}
\end{abstract}

\end{frontmatter}







\section{Introduction}
\label{sec:intro}

High-frequency radars (HFR) have been used for about half a century for the monitoring of coastal surface currents (\cite{Barrick_Science77}). As a secondary geophysical product, HFR also provide measurement of sea state parameters and wave spectra (e.g. \cite{wyatt_book21} for a recent review). However, this type of estimation is less accurate and less straightforward than the surface current estimation, {due to theoretical as well as practical reasons}. On the theoretical level, while the radial surface current only requires the measurement of the dominant first-order Bragg peaks, the wave spectrum is related to the side-band continuous components of the ocean Doppler spectrum surrounding the Bragg peaks. This second-order contribution has a complex analytical expression in the form of a nonlinear integral equation involving a coupling coefficient and quadratic combinations of the wave spectra at pairs of frequencies satisfying a resonance condition \cite{hasselmann_Nature71,barrick_tropo_1972,lipa_RS86}. Under some restrictive assumptions, \cite{barrick_RSE77,barrick_RS77} derived in $1977$ a method to extract the main sea state parameters from this second-order integral and accordingly, from the measured ocean Doppler spectrum. He showed that the significant wave height (SWH) and the {centro\"id wave frequency } can be expressed as simple ratios of weighted integrals involving the first- and second-order Doppler spectra. Soon after, linear and nonlinear inversion methods for retrieving the ocean wave spectrum from single {or dual} radar sites were developed in a series of papers (e.g. \cite{lipa_RS77,wyatt_IntJRemSens90,hisaki_RS96}). {Today, the inversion of the waves has reached a mature state and there are well-established techniques and commercial software for the determination of the ocean wave directional spectrum (see e.g. \cite{wyatt_book21} for a recent review). However, these inversion techniques require more expertise than that required to calculate the main wave parameters using Barrick's formulas. They have also physical as well as practical limitations (\cite{wyatt_JAOT00}) that restrict their range of applicability. The most critical issue relates to the quality of the radar data and, in particular, to the ability to measure the second-order Doppler spectrum with a sufficient signal-to-noise ratio.} Except for the high sea state and the highest radar frequencies, the second-order Doppler spectrum is usually several tens of dB below the level of the first-order peaks and reaches sooner the noise threshold when the range is increased. Hence, wave estimation methods based on the second-order Doppler analysis are confined to a restricted radar coverage in the near to middle range, depending on the radar frequency. Another issue is the clear identification of the second-order contribution and its separation from the first-order Bragg components, which can be ambiguous especially in the case of strong currents or in the absence of azimuthal resolution. For these reasons, {many authors have only considered the estimation of the bulk wave parameters using empirical methods (e.g. \cite{Essen_DHZ99,gurgel_JOE06,lopez_JTECH16}) or empirical adjustments of Barrick's method  \cite{heron_JAOT98,maresca_JGR80,ramos_JAOT09,alattabi_JTECH19,alattabi_JTECH21}}. Most of these methods necessitate the calculation of Doppler spectra that are resolved both in range and azimuth and can thus be associated with a reduced sea surface patch with constant sea state parameters. In the case of compact antenna arrays, the Doppler spectra are nondirectional, that is, they cannot be resolved in azimuth. This does not allow for the measurement of the wave frequency spectrum from a single station. Nevertheless, the $H_s$ can be estimated on range arcs adapting Barrick's original methods. The long-term performances of such measurements have been assessed by comparison with buoys and models and are quite variable depending on sea state, range, and direction of waves (see e.g. \cite{roarty_IEEE15,orasi_Meas18,saviano_Frontiers20} or \cite{lorente_OceanScience22} for a recent review).

In this paper, we propose a novel approach to sea state estimation based on the {time variation of the total radar signal, without applying the classical separation between first- and second-order components; instead, we exploit the amplitude modulation of the received voltage to infer the SWH and the peak wave frequency.} The idea of {using the time variation of the radar signal and bypassing the Doppler spectrum} to estimate the sea state parameters has been recently proposed in \cite{shahidi_GRSL20,shahidi_GRSL24} based on some mathematical approximations starting from the expression of the first-order electric field backscattered from a patch of the sea surface. It has been found that the SWH is simply proportional to the variance of the received voltage and that the peak frequency can be obtained from the same time series after integration over range. The method has been assessed with limited but promising experimental tests. Our approach, developed independently, is based on simple physical heuristic arguments and reaches a similar conclusion that the sea state parameters can be derived from the statistical properties of the received voltage and does not require any Doppler processing or identification of the second order contribution. However, our formulation is quite different in that it {does not reduce the signal to the sole contribution of the first-order Bragg term, but also includes higher-order terms through the hydrodynamic modulation of the backscattering. In addition, it is based on the calculation of the analytical and anti-analytical radar signal, a technique that allows us to isolate the contribution of long waves in the backscattered field.}

The paper serves a double objective. First, we test the performance of Barrick's method based on an annual data set of HF measurements with a radar located near Toulon. Its accuracy is very sensitive to some tuning parameters in its numerical implementation, which is part of the tacit know-how of the users and is rarely mentioned in the literature. A key step is the identification of the separation between the first- and second-order Doppler spectrum, for which we propose an empirical but explicit criterion. The second objective is to test the aforementioned new method based on the temporal properties of the backscattered signal. The HFR data set is described in section \ref{fig:data}, together with in-situ measurements and numerical models which are used for the assessment. The radar processing methods are presented in Section \ref{sec:methods} and a variety of statistical comparisons over one year of data is given in Section \ref{sec:results}  {while discussing the results obtained}.

\begin{figure}[h]
	\centering
	\includegraphics[scale=0.3]{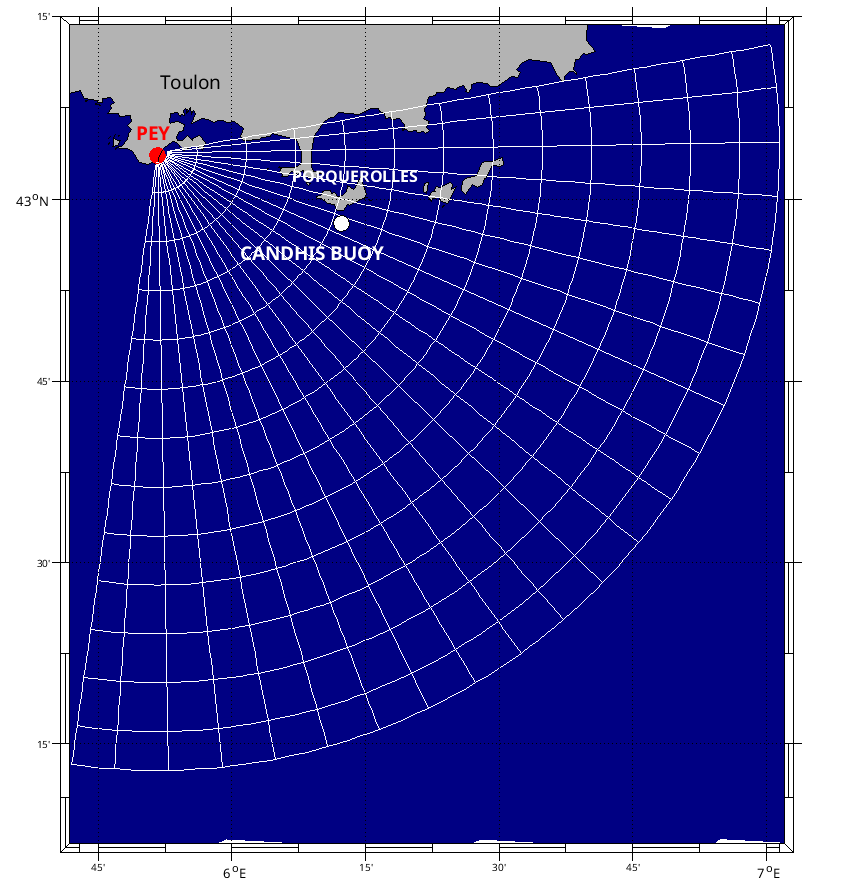}
        	\caption{Location of HFR coverage in the Mediterranean region of Toulon (France). The red dot marks the location of Fort Peyras, the blue dot that of the CANDHIS buoy, and the white lines represent {5 range units} (every 7.5 km) and {5 azimuth units} (every $5\rm{^{o}}$) of the radar coverage.}\label{fig:Figure1_HFRlocation}

\end{figure}

\begin{figure*}[h]
  \centering
  \includegraphics[scale=0.5]{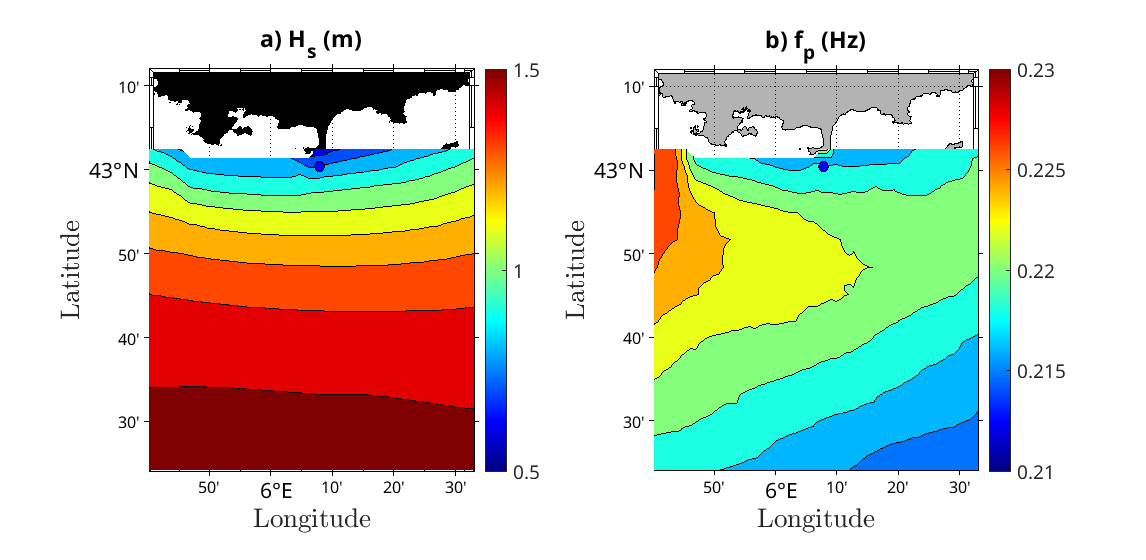}
        \caption{{Annual mean value  of a) the SWH (in meter) ; b) the peak wave frequency in the radar coverage according to the WWIII model. The blue dot marks the position of the CANDHIS buoy.}}	\label{fig:Figure2_Hs_WWIII}

\end{figure*}

\section{Data}\label{fig:data}
\subsection{High Frequency Radar data}
The Toulon HFR network, developed by WERA Helzel Messtechnik, consists of a combination of two transmitters and two receivers located at three different sites. On the island of Porquerolles, $27$ km southeast of Toulon, there is a stand-alone transmitter with a single omnidirectional antenna covering a wide area to the south. Cap Bénat, located $35$ km east of Toulon, hosts the first set of receivers, consisting of a linear array of $12$ active antennas oriented $70$ degrees counterclockwise from the north, with an antenna separation of $0.45\lambda$. The second set of transmitters and receivers is located at Fort Peyras, about $8$ kilometers southwest of Toulon, with a linear array of $12$ passive antennas oriented north-south, also with a spacing of $0.45\lambda$. This configuration of transmitters and receivers produces three bistatic pairs and one monostatic pair \cite{guerin_Radar19,dumas_OD20,dumas_JTECH22}.

In this study, we use the monostatic signals collected by the Fort Peyras site (figure \ref{fig:Figure1_HFRlocation}) during $2020$, a year for which data availability was over $91\%$. This radar operates in a $16.15$ MHz center frequency band with an assigned bandwidth of $100$ kHz. Depending on weather conditions, it can cover a range of $60$ km with a range resolution of $1.5$ km and an angular sector of about $110\rm{^{o}}$.
The data consist of the complex I/Q voltage recorded on the set of $12$ receive antennas and sorted in range after a preliminary Doppler processing in the fast time. The sampling rate in the slow time is $0.26$ seconds and the time series are acquired in continuous blocks of $4096$ values (lasting about $18$ min) every $20$ min, which corresponds to the standard ``SORT file'' output format of WERA HFRr. The data are further processed in azimuth by applying the classical beamforming algorithm ({with steering steps of one degree}) to the $12$ channels of the antenna array. This leads to a complex discrete time series $s(t)$ for each {radar cell of size $1.5$ km$\times 1$ degree} and each $20$ min sequence. {In the following, each radar cell is labeled by a range number ($1$ to $64$ starting from the radar) and an azimuth number ($1$ to $110$ starting from the westernmost corner).}

\subsection{Wave data}

{We recall the standard definitions of the SWH as the integral of the wave frequency spectrum $S(f)$:

  \begin{equation}\label{eq:defHs}
    H_s=4\left(\int_{0}^{\infty}S(f)df\right)^{1/2}
  \end{equation}

which can also be obtained as four times the RMS elevation of the free surface. The peak or dominant frequency is defined as:
 \begin{equation}\label{eq:deffp}
   f_p=\mathrm{Arg Max} S(f)
   \end{equation}

and the mean or centro\"id wave frequency $f_c$ is defined with the first moment of the wave frequency spectrum $S(f)$:
  
\begin{equation}\label{eq:deffc}
f_c=\frac{\int_{0}^{\infty}fS(f)df}{\int_{0}^{\infty}S(f)df}.
\end{equation}

As we will see, the peak frequency $f_{p}$ turns out to be more relevant than the centroïd wave frequency ${f_c}$ for the comparisons with radar-based estimates, at least in the case of a wind-wave sea with non distinct swell component, as it is in general the case in the area of Toulon.

For the year $2020$ we used as ground truth a time series of wave parameters collected by a CANDHIS (Centre d'Archiveage National de Données de Houle In-Situ) buoy located in the vicinity of Porquerolles island, at about $32$ km from the radar ($42\rm{^{o}}58.00'$N, $6\rm{^{o}}12.29'$E, see Fig. \ref{fig:Figure1_HFRlocation}). These data are provided by the CEREMA (\textit{Climat et Territoires de Demain)} at a temporal resolution of $30$ minutes and they are available in \cite{CANDHIS}. They cover the whole year 2020 with the exception of a 83 days gaps between May, 5 and July, 27. They provide the $H_s$ calculated with four times the RMS elevation and the peak frequency $f_p$.}

\begin{figure}[h]
	\centering
	\includegraphics[scale=0.6]{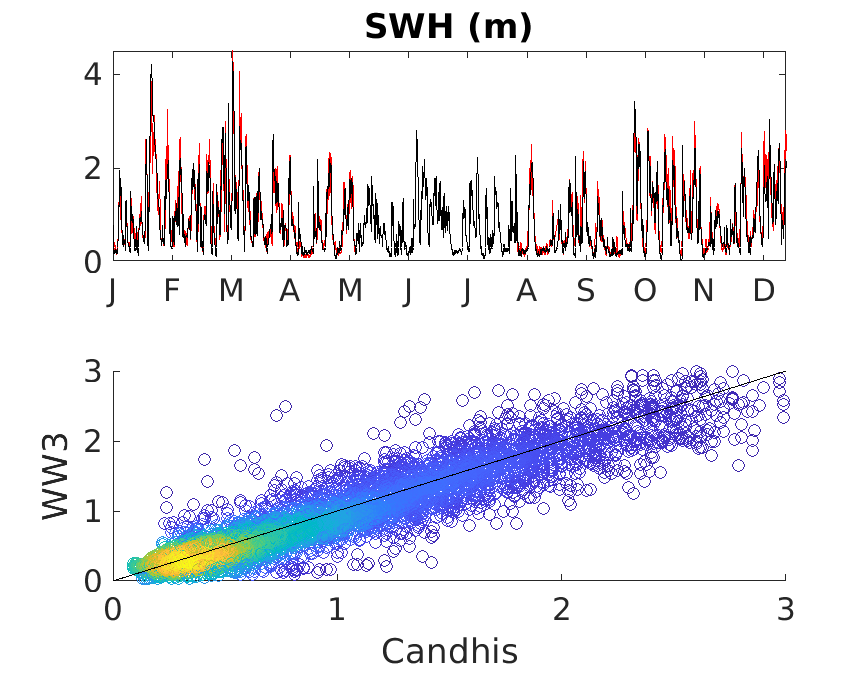}
        \caption{{Time series (top) and scatterplot (bottom) of the SWH (in meter) over the year 2020 at the CANDHIS location according to the WWIII model (black line) and according to the buoy (red lines).}}
	\label{fig:Figure3_Hs_WWIII_Candhis}
\end{figure}

\begin{figure}[h]
	\centering
	\includegraphics[scale=0.6]{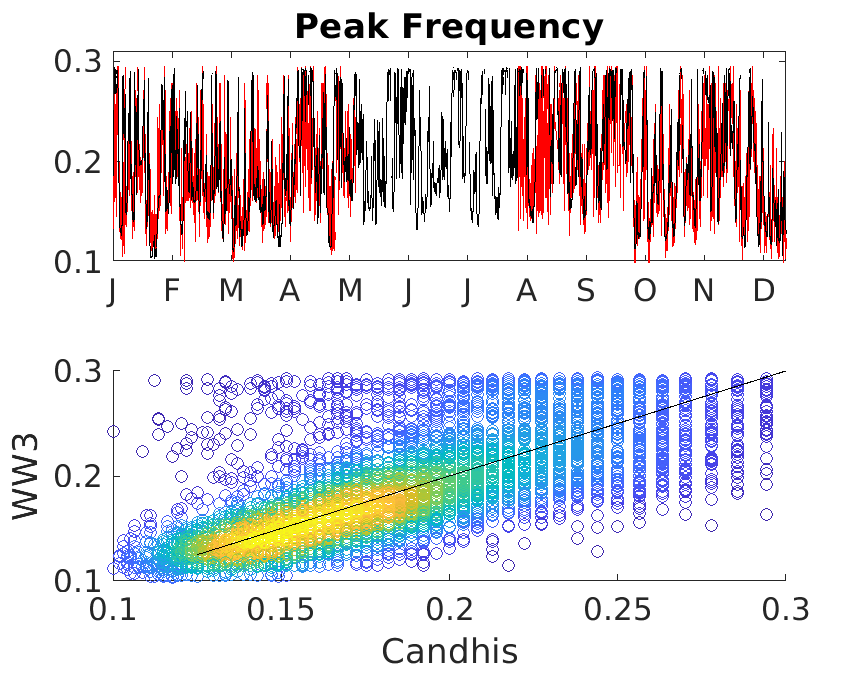}
        \caption{{Same as Figure \ref{fig:Figure3_Hs_WWIII_Candhis} for the peak wave frequency}}
	\label{fig:Figure3_fp_WWIII_Candhis}
\end{figure}

{During the same year, the results of the Wavewatch III model (WWIII) at an hourly time resolution were available over the study area on a 3-level nested grid (with a spatial resolution of $800$m in the radar \col{area}) forced by $10$-m wind fields obtained from the non-hydrostatic mesoscale Weather Research and Forecast model \citep[for more details consult][]{LLAAOR2022}. The SWH and the mean wave frequency are estimated through the spectral moments (\ref{eq:defHs}) and the peak wave frequency $f_p$ is also provided.  The annual mean values of the SWH and the peak wave frequency in the coastal region of Toulon according to the WWIII model are shown in figure \ref{fig:Figure2_Hs_WWIII}. As seen, the typical SWH in the radar aera is of the order of $1$ m while the typical value of the peak frequency is about $0.22$ s.  
While the ground truth in the vicinity of the radar area is limited to the single location of the CANDHIS buoy, it is possible to use the WWIII model to complement the analysis. To check the accuracy of the WWIII in the radar area, we compared the time series of $H_s$ and $f_p$ at the buoy location to the actual measurements. The point by point comparison is shown in the scatterplots of figures \ref{fig:Figure3_Hs_WWIII_Candhis}-\ref{fig:Figure3_fp_WWIII_Candhis}. The Root Mean Square Difference (RMSD) of SWH is $25$ cm, which is about $25\%$ of its mean value ($0.96$ m) while the RMSD of the peak wave frequency is $0.037$, which is about $20\%$ of its mean value ($0.19$ Hz). The Pearson correlation coefficient is higher for the SWH ($0.94$) than for the peak wave frequency ($0.70$). The statistical parameters to compare the different time series are summarized in Table \ref{tab:CANDHIS_WW3}.}

\begin{table}[h]
  \caption{{Mean wave parameters from the CANDHIS buoy and estimation of the  WWIII/CANDHIS difference over the year $2020$.}}

\footnotesize{
  \begin{tabular}{|c|c|c|c|c|c|}
    \hline
 &   CANDHIS &&& WWIII/CANDHIS&\\
    \hline
 & mean & RMS   & mean bias & RMSD          &  R   \\
    \hline
$H_s (m)$ &$0.96$& $0.71$ & $-0.03$& $0.25$  & $0.94$   \\ \hline
    $f_p$ (Hz) & $0.19$ & $0.044$ & $0.004$ & $0.037$ & $0.70$  \\
\hline
\end{tabular}
}
\label{tab:CANDHIS_WW3}.
\end{table}

\section{Material and Methods}\label{sec:methods}

Two different methods {have been} used to estimate the ocean waves, $H_s$ and $f_p$, from HFR signals. The first is the classical Second Order Doppler Spectrum method, first introduced by \cite{barrick_RS77,barrick_RSE77}, { while the second is an original method that uses only the time modulation of Bragg wave amplitude. In the following we will refer to these methods simply as ``Barrick's method'' and the ``Modulation method''.}

\subsection{Second Order Doppler Spectra Method}\label{subsec:SODS}

\begin{figure}[h]
	\centering
	\includegraphics[scale=0.35]{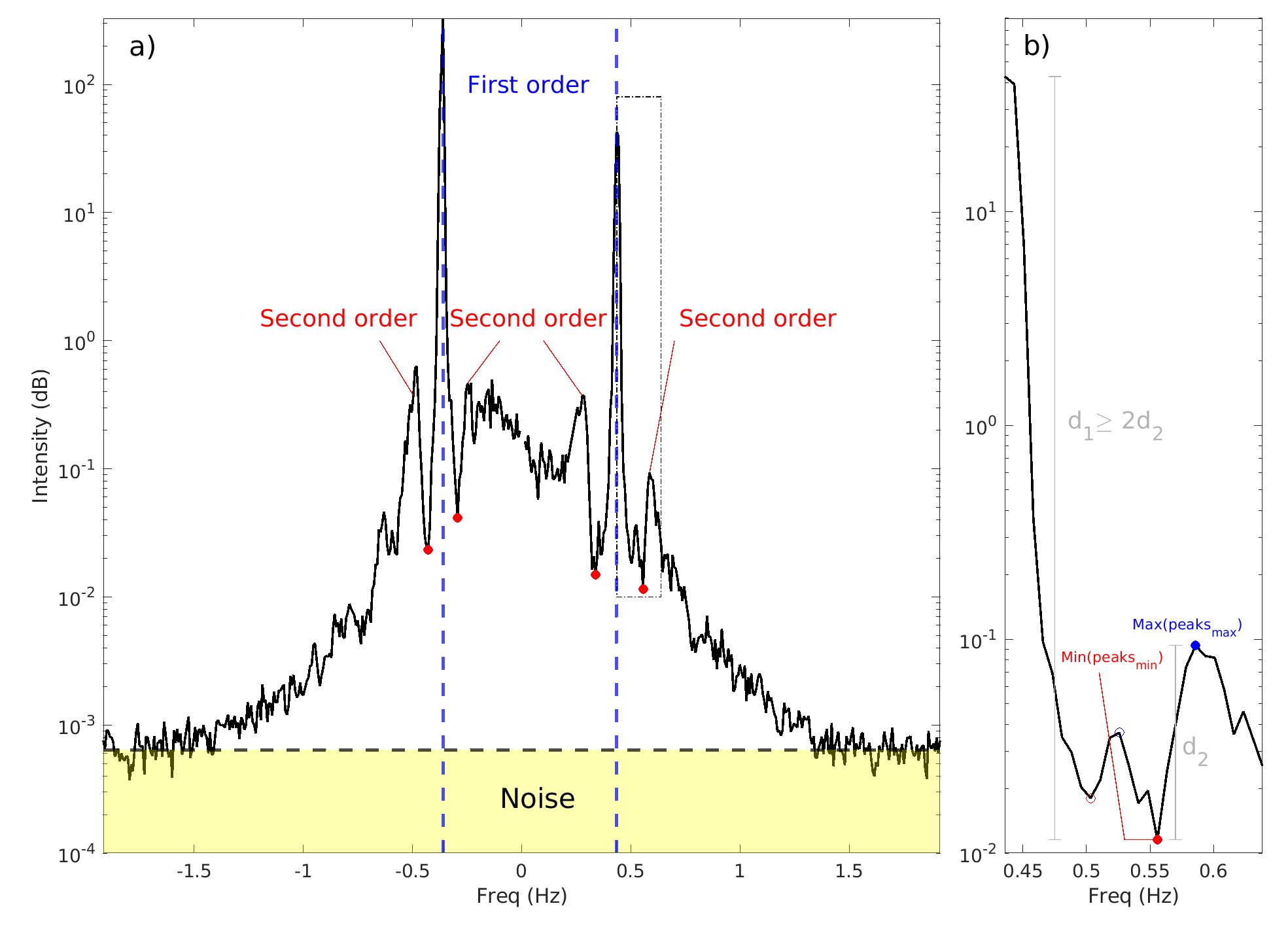}
        	\caption{a) Doppler spectrum for March $8$, $202$0 at $0$h at azimuth $40$ and range $8$ (see figure \ref{fig:Figure1_HFRlocation}, b). The dashed blue lines mark the first-order peaks. The second-order peaks are also marked in the figure, selecting the separation points between the first and second order with filled red dots. The noise threshold is marked with a dashed line and yellow shading. b) Zoom in on the right side of the positive first-order peak of the Doppler spectrum. The minimum/maximum peaks (red/blue points) when delimiting the first order from the second order of the spectrum are shown.}

	\label{fig_Barrick}
\end{figure}

\begin{figure}[h]
	\centering
	\includegraphics[scale=0.25]{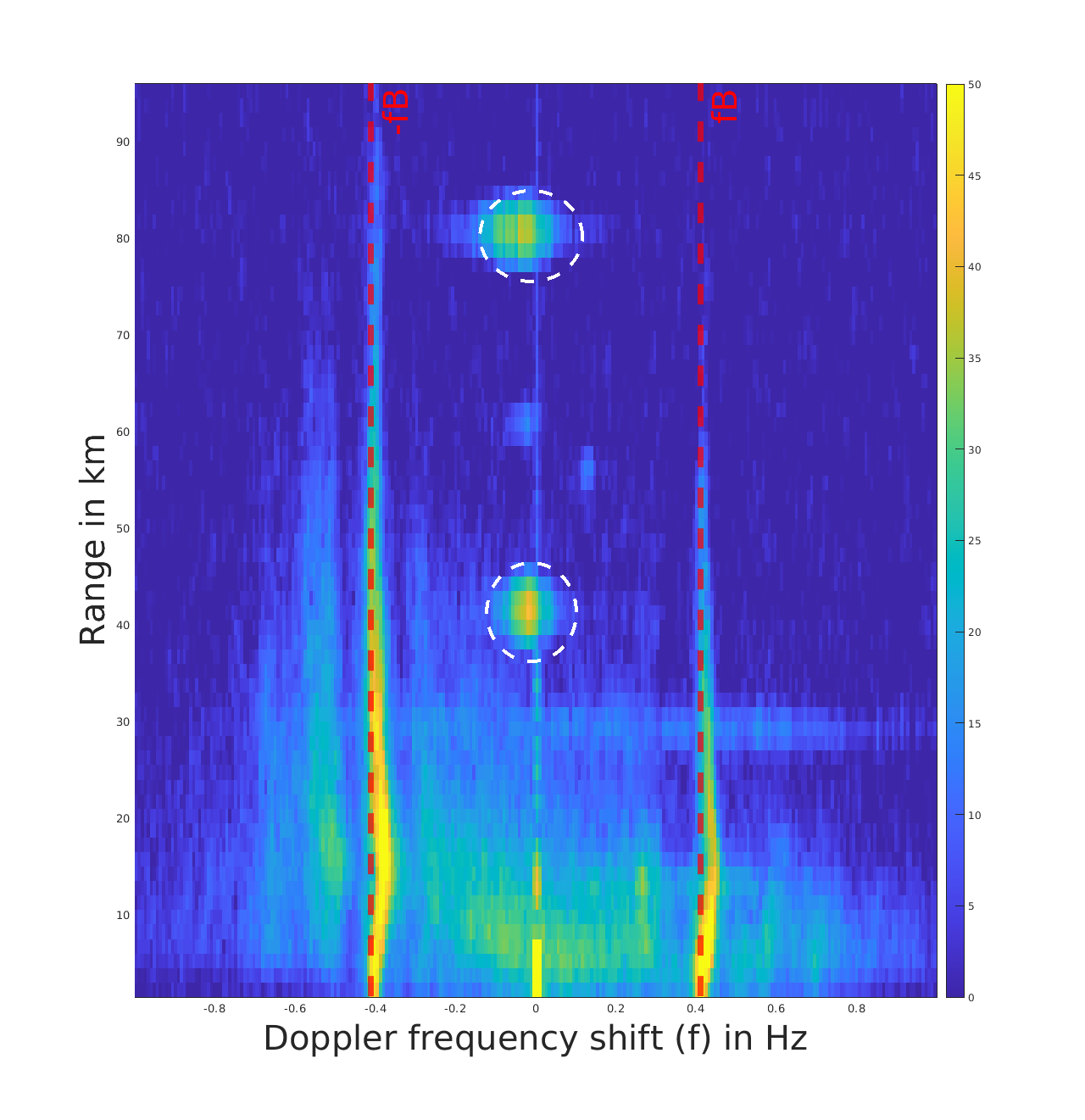}
        \caption{Range Doppler spectrum (in dB) for March $8$, $202$0 at $0$h at azimuth $40$ (see figure \ref{fig:Figure1_HFRlocation}, b). The dashed red lines mark the Bragg frequency and the dashed white circles mark the few ranges where the zero-Doppler region is corrupted by power grid artifacts ($\sim 42$ and $82$ km). }
	\label{fig:rangedop}
\end{figure}

The classical method for estimating $H_s$ and $f_p$ from the HFR sea echo originates from the seminal work of \cite{barrick_RS77,barrick_RSE77} and involves straightforward calculations based on the ocean Doppler spectrum.
The hourly range-azimuth resolved Doppler spectra are classically obtained by forming the power spectrum of the radar signal:

\begin{equation}
  \sigma(f)=\langle \abs{\hat s(f)}^2\rangle,
\end{equation}

where $\hat s(f)$ is computed in the standard way on a discrete set of frequencies using a Fast Fourier Transform of the time series with a Hamming attenuation window and an integration time of about $2$ min. To reduce the noise variance, a further incoherent summation is applied by taking a median average of the sample spectra $\abs{\hat s(f)}^2$ over $15$ overlapping intervals within the $18$ min duration of each available SORT file (3 per hour). This will result in an average Doppler spectrum with a bandwidth of $4$ Hz (corresponding to a sampling rate of $0.26$ sec) and a frequency resolution of $\Delta f=1/T\simeq 0.008$ Hz. An example of such a Doppler spectrum is shown in figure \ref{fig_Barrick}, a) for the $8$ range cell ($12$ km) and azimuth $40$ for March $8$, $2020$. The corresponding range Doppler spectrum along the same direction is shown in figure \ref{fig:rangedop}. In this example, the strongest first-order Bragg line extends to about $90$ km, while the strongest second-order sideband is visible up to $65$ km away from the radar. The dashed white circles highlight a power grid artifact ($50$ Hz harmonics) that contaminates the low Doppler frequency content of the radar signal at some ranges; this electrical artifact is particularly pronounced at the Fort Peyras site due to the imperfect power supply at this location. It does not affect the estimation of the radial current from the first-order Bragg lines but must be treated with caution when dealing with the full ocean Doppler spectrum.

To determine $H_s$, the ratio of the weighted integral of the second-order Doppler spectrum to the energy associated with the first-order peaks is analyzed (see figure \ref{fig_Barrick},a).

\begin{equation}\label{eq:Hs}
H_s^2=\frac{32 \int_{-\infty}^{\infty} \sigma_2(f)/{\cal W}({f/f_B}) df}{k_0^2 \int_{-\infty}^{\infty} \sigma_1(f) df},
\end{equation}

where $\sigma_1$ and $\sigma_2$ are the first and second order Doppler spectra, respectively; ${\cal W}$ is the weighting function defined in \cite{barrick_RSE77}, $k_0$ is the scalar radio wavenumber, and

\begin{equation}
  f_B=\sqrt{\frac{g}{\pi\lambda}}
\end{equation}

is the Bragg frequency and $g=9.81\ m.s^{-2}$ is the gravitational constant. The mean wave frequency can also be determined, in principles, from a different ratio of integrals based solely on the second-order Doppler spectrum:

\begin{equation}\label{eq:fp}
f_c = \frac{\int_{0, f_B}^{f_B, \infty} \lvert f-f_B\rvert \sigma_2(f)/{\cal W}(_{f/f_B}) df}{ \int_{0, f_B}^{f_B, \infty} \sigma_2(f)/{\cal W}({f/f_B}) df},
\end{equation}

where the bounds of the integrals can be either $0$ and $f_B$ or $f_B$ and $+\infty$, assuming that the positive frequency sidebands are the strongest (otherwise one should take the symmetric frequency intervals on the negative side). In this study, the maximum value of the integral of both intervals (from $0$ to $f_B$ and from $f_B$ to $+\infty$) is chosen for each time. {In reality, the maximum available Doppler frequency for the second-order Doppler spectrum extends at best to 3 or 4 times the Bragg frequency (i.e. about $1.5$ Hz at our operating radar frequency $16.15$ MHz), so that the above integral is actually closer to the peak frequency than to the mean wave frequency.}


The main challenge of Barrick's method is to effectively discriminate between first and second-order spectra and to accurately distinguish signal from noise. To overcome this, specific techniques have been developed that allow the precise setting of separation thresholds and the establishment of control criteria for integration regions. The effectiveness of this technique is limited by its dependence on the second-order Doppler spectrum, which restricts its application to shorter distances where this spectrum remains well above the noise level. {Due to this greater sensitivity to noise, it is important to subtract the mean noise level from the second-order Doppler spectrum before performing the integration in (\ref{eq:Hs}) and (\ref{eq:fp}).} \col{The first-order Bragg peaks were identified in a classical way by seaching the maxima of the Doppler spectrum inside the first-order Bragg region, whose maximal excursion is determined by the maximal value of the radial current in the area.} To determine the exact point of transition from first to second-order spectra, and thus avoid false boundaries, the following method is used: positive peaks (marked with a blue circle in figure \ref{fig_Barrick}, b) and negative peaks (marked with red circles in figure \ref{fig_Barrick}, b) are selected. The minimum negative peak (marked with a solid red circle in figure \ref{fig_Barrick}, b) and the maximum positive peak at frequencies higher than the minimum negative peak (marked with a solid blue circle in figure \ref{fig_Barrick}, b) are identified. We impose the condition that the intensity difference between the first-order peak and the minimum negative peak ($d_1$) must be at least twice the difference between the minimum negative peak and the maximum positive peak ($d_2$). If this condition is not met, the selected minimum negative peak is discarded and the process is repeated with another negative peak initially identified. {This threshold value of $2$ is purely empirical and has been found to be effective in removing satellite peaks near the first-order Bragg peak that could be misinterpreted as second-order peaks. Although these peaks are located at very close frequencies, the resulting error in calculating the integral of the second-order Doppler spectrum can be large, resulting in an inaccurate estimate of $H_s$.}

At the same moment, the noise level is determined as the average intensity for frequencies higher than $1.75$ Hz (see figure \ref{fig_Barrick}, a). {In spite of these tests, the correct identification of the second-order Doppler spectrum may sometimes fail and lead to erroneous estimates of the wave parameters. However, a significant proportion of these irrelevant values can be eliminated by a posteriori quality checks such as the removal of outliers in the time series of estimated wave parameters.}

\subsection{{The Bragg Wave Modulation Method}\label{subsec:BWMM}}

\begin{figure}[h]
	\centering
	\includegraphics[scale=0.3]{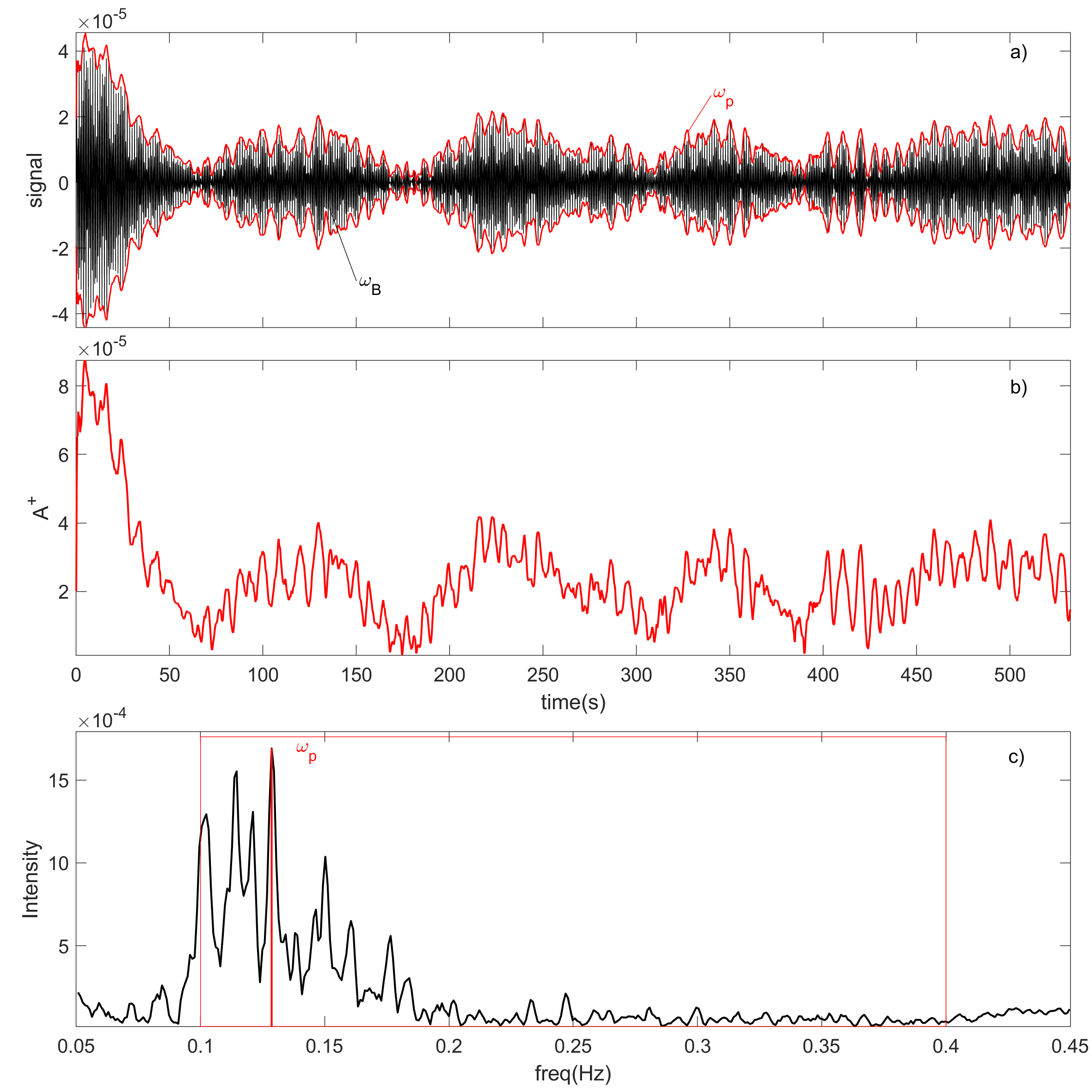}
        	\caption{a) Real part of the received radar signal complex time series (black solid line) for the range $10$ ($15$ km) and azimuth $75$ on September $27$, $2020$, at $23$h  where $H_s$ and $T_p$ from CANDHIS are $2.93\ m$ m and $7.5\ s$, respectively. {The red solid line is the signal envelope, obtained with the modulus (and its opposite value) of the analytical signal.} b) Modulus of the analytical signal for the same {time}. c) Sum of the analytical and anti-analytical power spectra ($\sigma_{\abs{s_a^+}}(f)+\sigma_{\abs{s_a^-}}(f)$) for the same time. The common wave frequencies in the Mediterranean Sea are marked with a red rectangle where the maximum power spectrum is related to peak wave frequency ($f_p$)}
	\label{Figure4_signalanal}
\end{figure}

{It is well known that the backscatter from short waves is modulated by the slow variation of long waves through nonlinear hydrodynamical interaction mechanisms. The variation of the received power from the resonant short wave is usually represented by the so-called Modulation Transfer Function (MTF). Even though the concept of MTF has been mostly defined in the context of microwave scattering (e.g. \cite{keller_RS75,keller_JGR94}), it also applies to HFR whenever there are waves in the ocean that are much longer than the Bragg wave. Here we adapt the concept of MTF to the simplified cased where there is only one long wave (say, the peak wave). Its frequency and wavenumber are assumed to be much smaller than those of the Bragg frequency ($f_p<<f_B$, $k_p<<k_B$) while its amplitude is much larger ($A_p>>A_B$) so we can assume a simple Two-Scale model. Considering the hydrodynamic interaction of the lowest order in long-wave steepness, we have that both the amplitude and wavenumber of the Bragg wave are modulated by the long-wave orbital cycle $\Phi$ \citep{Longuet_JFM60}:

\begin{equation}\label{eq:a_B}
\tilde a_B(t)=a_B(1+k_pa_p\cos\Phi) 
\end{equation}

\begin{equation}\label{eq:k_B}
\tilde k_B(t)=k_B(1+k_pa_p\cos\Phi),
\end{equation}

Assuming that the dominant part of the sea echo is due to first-order Bragg scattering and allowing the Bragg wave to propagate in both directions, a similar amplitude modulation will result for the complex radar signal:

\begin{equation}\label{eqsignal}
  s(t) \sim A_B^+(t) e^{2\eye\pi(\tilde f_B+f_u)t}+ A_B^-(t) e^{-2\eye\pi(\tilde f_B-f_u)t}
\end{equation}

where $t$ is time; $A^\pm_B$ are complex amplitudes of the form (\ref{eq:a_B}) with random phases; $\tilde f_B$ is the modulated Bragg frequency and $f_u=2 u/\lambda$ is the additional frequency shift due to the radial surface current $u$. By hypothesis, the variance of the amplitude is proportional to the variance of the long wave steepness:

\begin{equation}
\var{\abs{A_B^\pm}}\sim \var {k_pa_p^\pm}
\end{equation}

and thus the same holds for the SWH:

\begin{equation}\label{eq:HsvarR}
    H_s^2\sim \var{\abs{A_B^\pm}}
\end{equation}

The  complex Bragg wave amplitudes $A^\pm_B$ are generally different but if we assume that they are uncorrelated due to random phase shifts between the approaching and receding waves, we can simply sum their contribution to the SWH:

\begin{equation}\label{eq:fpvarR}
    H_s^2\sim \var{\abs{A_B^+}}+\var{\abs{A_B^-}}
\end{equation}

}
Note that the oscillation factor due to long wave modulation ``survives'' the radar cell averaging process because the wavelength of the longest waves is not negligible compared to the radar range resolution. For typical scales of $75$ m for the former and $1.5$ km for the latter, about $20$ long waves pass through a given radar cell, so that the mean surface level within that patch oscillates in phase with the long wave with a damping factor of about $1/20$. The amplitude of this damped oscillation remains large compared to the amplitude of the Bragg wavelength (because the ratio of the peak wavelength to the Bragg wavelength is $75/9.3\sim 8$ and the ratio of the corresponding amplitude is $\sim 8^2$ for a $k^{-4}$ wave spectrum). In addition, there are also wave groups of comparable or larger size to the radar cell. These groups are responsible for very low frequency modulation of the radar signal. Since both the peak wave frequency $f_p$ and the current-induced frequency shift $f_u$ are much smaller than the Bragg frequency $f_B$, we can rewrite eq. \eqref{eqsignal} as the sum of an analytical signal $s_a^+$ (respectively, anti-analytical signal $s_{a}^-$) that contains only the positive frequency side (respectively, negative frequency side) of the Bragg peaks:{

\begin{equation}\label{eqsasaa}
s(t) = s_a^+(t) +s_{a}^-(t)
\end{equation}

where

\begin{equation}\label{sa}
s_a^\pm(t)\sim A_B^+(t) e^{\pm 2\eye\pi(\tilde f_B\pm f_u)t}
\end{equation}

It follows that the SWH can be extracted from the complex radar signal by taking the absolute value of its analytical and anti-analytical components:

\begin{equation}\label{eqHs2}
H_s^{2} =C (\var{\abs{s_{a}}}+\var{\abs{s_{aa}}}
\end{equation}

where $C$ is some calibration factor that must be set for each range and azimuth. In the following, this factor has been calculated for each range cell and azimuth by adjusting the mean value of the right-hand side in (\ref{eqHs2}) $H_s$ for the whole $2020$ year to the mean value of the $H_s$ obtained with the WWIII model.
}

{According to the heuristic Two-Scale model underlying the modulation approach, the slow variation in amplitude $\abs{A^\pm_B(t)}$ follows the cycle of the long wave. However, it is clear than in general there is a continuum of scales between the peak wave and other long waves and we expect the slow amplitude to vary according to the peak wave frequency ($f_p$). It can be identified at the maximum of the associated power spectra $\sigma_{\abs{s_a^\pm}}$ (which is defined as the mean square modulus of the Fourier transform of $\abs{s_a^\pm}$):}

\begin{equation}
f_p = \mathrm{Argmax}(\sigma_{\abs{s_a^+}}(f)+\sigma_{\abs{s_a^-}}(f))
\end{equation}

{In practice we calculate the power spectra from $18$ min time series (corresponding to the duration of a WERA SORT file) by applying a Fourier transform on the entire sample and smoothing the spectrum with a $10$ point moving average over frequencies.} Figure \ref{Figure4_signalanal} a) shows an example of the complex radar signal time series recorded on the antenna receiving array after beamforming processing. A slow amplitude modulation is visible, marked by the red lines. Part of this modulation is due to the beat frequency caused by the shift of the surface current from the Bragg frequency (see \cite{guerin_OM18}) and another part is due to the long wave groups ({eqs. (\ref{sa})}). This last term can be isolated by taking the modulus of the analytical/anti-analytical signal (see figure \ref{Figure4_signalanal}, b)), which has the effect of canceling out the Bragg component and the current-induced frequency shift. The peak wave frequency is sought at the maximum of the corresponding power spectrum (figure \ref{Figure4_signalanal}, c)) in the range $0.1$ to $0.4$ Hz, which contains the peak wave frequencies for the Mediterranean Sea. Only the positive part of the power spectrum is shown as it is symmetric. {Note on figure  \ref{Figure4_signalanal}, b) the very slow modulation at a typical time scale of $100$ sec. This corresponds to very low frequencies (about $0.01$ Hz) that fall far outside of the range of wind-wave gravity waves and could correspond to other phenomena such as infragravity waves.}

\section{Results and Discussion}\label{sec:results}
{The hourly SWH and peak wave frequency were calculated from the radar data using the above procedure for each radar cell. Due to the azimuthal step of beamforming (1 degree), which is much smaller than the actual angular resolution obtained with a 12 antenna array ($~10$ degrees), the outputs of neighbouring radar cells are highly correlated and we therefore performed a moving average over 5 degrees. Estimations of wave parameters that are outside the permissible range of values expected in the coastal region of Toulon (i.e, such that $H_s>5$ m and $f_p<0.1$) were eliminated. In addition, an upper threshold was set at $0.35$ Hz for the peak wave frequency calculated with the modulation method, due to the necessary separation of long wave scales. Further outliers were eliminated by removing the nonphysical jumps in the time series (using a threshold of 50 cm and 0.1 sec, respectively, for the permissible change of $H_s$ and $f_p$ over one hour). }
\subsection{Statistical performances at the buoy location}\label{sec:perfCANDHIS}
           {The CANDHIS buoy is located at a distance of about 32 km from the radar. Although this relatively large distance is not optimal for estimating the wave parameters from the HFR data, the buoy data can be used to assess the performance of the two proposed methods.   Figures \ref{Figure7_timeseriesHs_atCandhis} and \ref{Figure7_timeseriesfp_atCandhis} show the time series (re-centered over one month for better visibility) for the SWH and peak wave frequency, respectively, using either the HFR-based estimates, the CANDHIS records or the WWIII model. \col{It appears that Barrick's method gives very consistent estimates compared to the buoy measurements after correction of the $H_s$ estimate with an overall $0.58$ scaling factor. The Bragg wave modulation method, however, is less reliable in estimating the wave parameters, althrough its estimates clearly follow the observed trends of variations.} 

           {To compare the methods in more detail, the radar/buoy scatterplots are shown in figures \ref{Figure8_scatterplotBarrick_atCandhis} and \ref{Figure8_scatterplotModulation_atCandhis} for the annual time series where simultaneous data are available. Along with these plots we calculated the usual statistical parameters, namely the RMSD and the Pearson correlation coefficient (R) (see Table \ref{table_perf}). It confirms a better correlation of Barrick's method with the buoy measurements ($R=0.87$ and $R=0.62$ for the SWH and peak wave frequency, respectively ) while the estimates obtained with the Bragg Wave Modulation Method are more scattered ($R=0.43$ and $R=0.38$, respectively). It also shows that Barrick's method has a tendency to overestimate both $H_s$ and $f_p$, \col{calling for an extra corrective factor. Such empirical scaling factor has been used by several authors \citep{maresca_JGR80,heron_comparison_1998,ramos_JAOT09} to adjust the radar estimates of $H_s$. One invoked reason for applying such correction is that the weighting function used in Barrick's formulas (\ref{eq:Hs}) and (\ref{eq:fp}) favors waves that propagate perpendicular to the beam direction \cite{wyatt_book21}.} We found that a scaling factor of about $0.58$ significantly improves the agreement with the buoy and reduces the RMSD from $0.48$ m to $0.29$ m, which is close to the RMSD obtained with the WWIII model ($0.25$ m). Note that this scaling factor is very consistent with the values found in  \citep{maresca_JGR80} ($0.55$) \cite{ramos_JAOT09} ($0.58$). }

\begin{figure}[h]
\hspace{0cm}\includegraphics[scale=0.15]{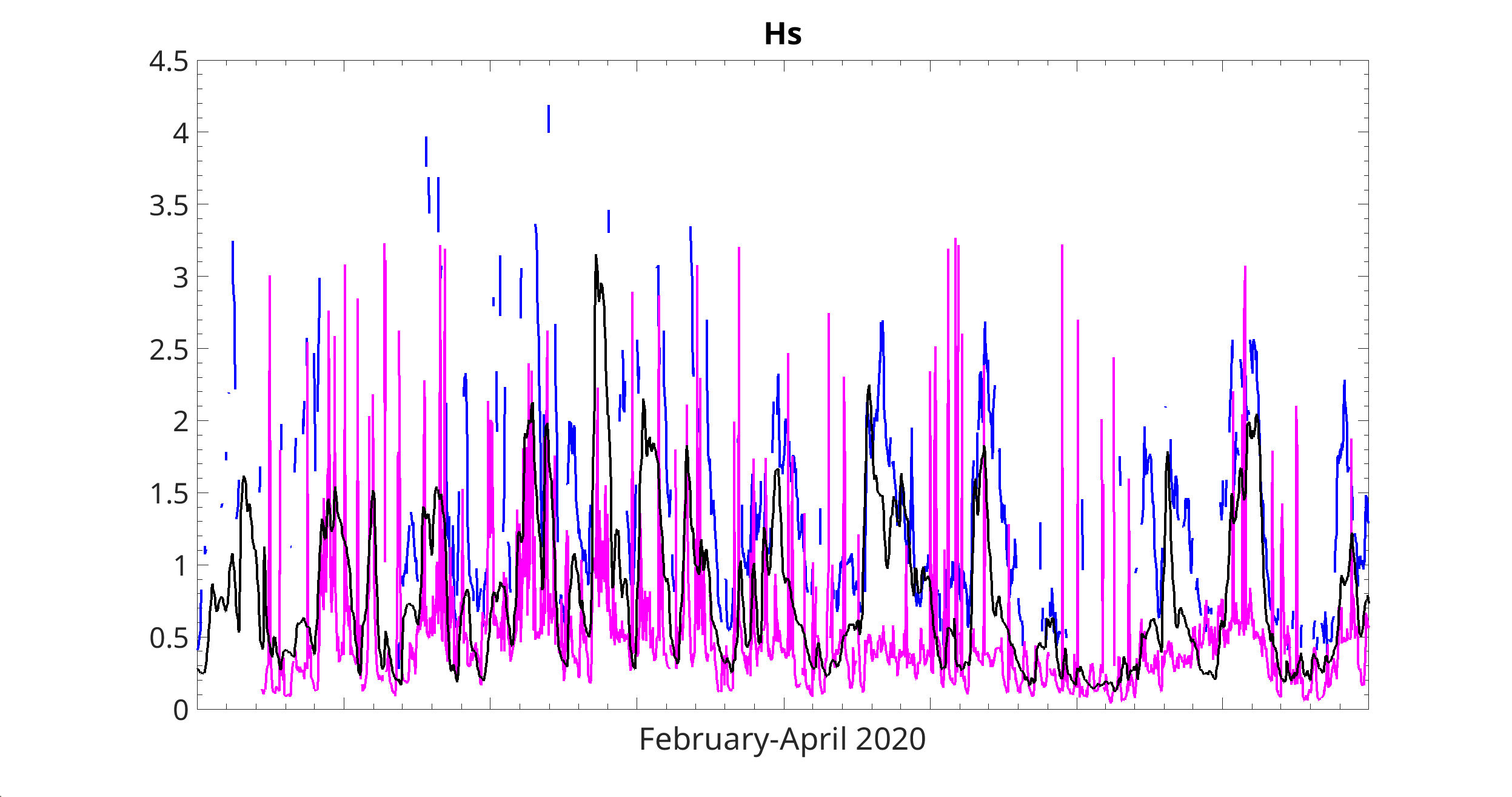}
  	\caption{{Time series of $H_s$ (m) from February 1st to April 30, 2020 at the location of the CANDHIS buoy, as estimated by Barrick's method (blue lines) or the Modulation method (magenta lines). The buoy measurements are shown with black solid lines.}}
	\label{Figure7_timeseriesHs_atCandhis}
\end{figure}

  \begin{figure}[h]
 \hspace{-1cm}\includegraphics[scale=0.15]{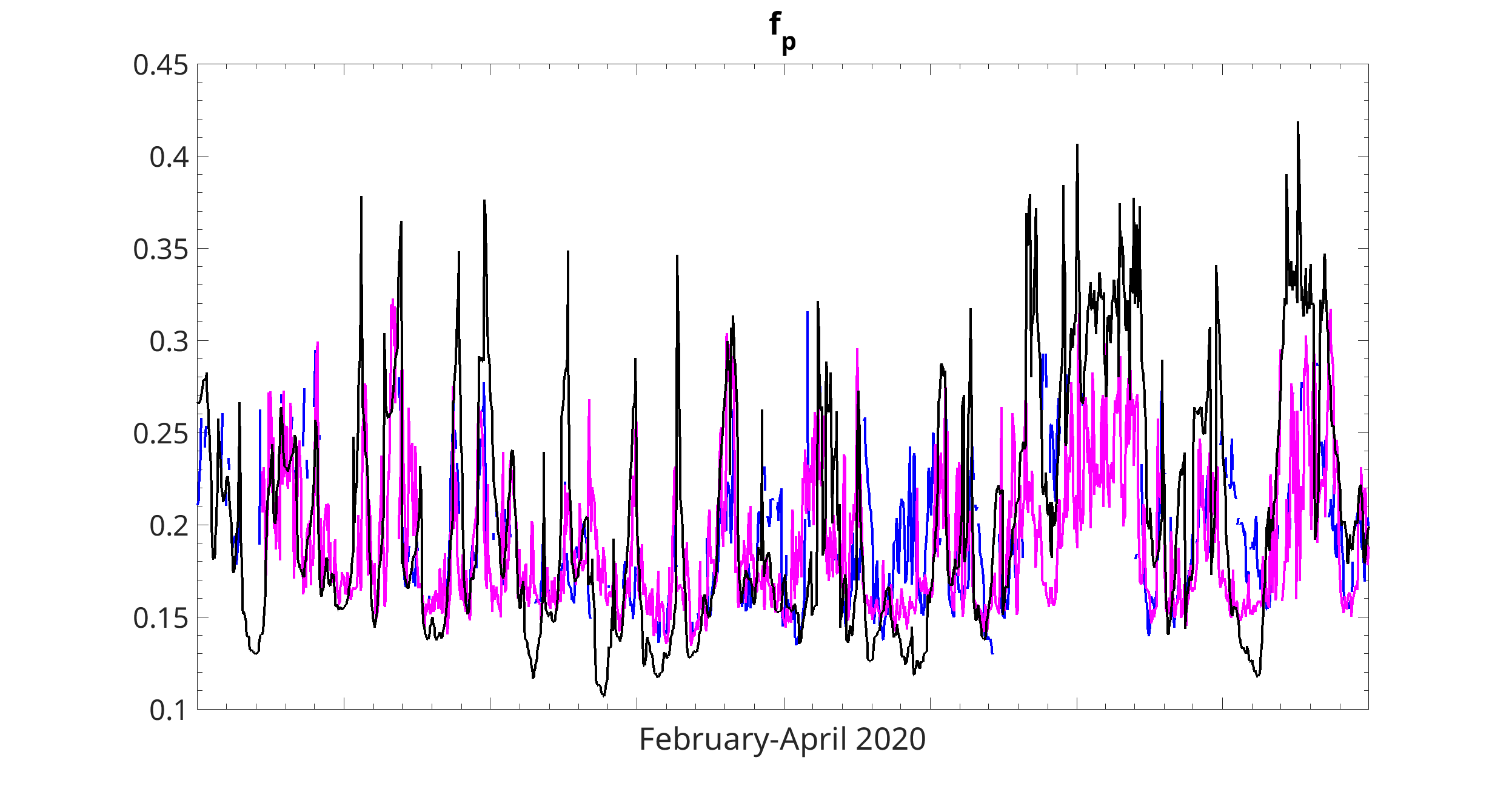}
	\caption{Same as Figure \ref{Figure7_timeseriesHs_atCandhis} for the peak wave frequency $fp$ (Hz).}
	\label{Figure7_timeseriesfp_atCandhis}
\end{figure}

  \begin{table*}[h]
    \centering
    \caption{{Mean bias ($\mu$), RMSD ($\sigma$) and Pearson correlation coefficient ($R$) of Barrick's method (B), the Modulation method (M) and the WWIII model (W) for the estimation of $H_s$ (in meter) and peak wave frequency $f_p$ (in Hz) over the year $2020$ by comparison with the CANDHIS buoy measurements (range number $21$ and azimuth number $78$). The RMSD of $H_s$ following from Barrick's estimate is given without ($0.48$ m) and with ($0.29$ m) the scaling factor $0.58$.}}

\footnotesize{
  \begin{tabular}{|c|c|c|c||c|c|c||c|c|c|c|c|}
    \hline
&{$\mu$}&&    & {$\sigma$} &         &     &{$R$} &     &      \\
    \hline  
    & B & M & W &  B & M & W& B & M & W \\ \hline
    $H_s$ &$0.48$ & $-0.11$ &$-0.03$ &  $0.48$  & $0.65$ & $0.25$ & $0.87$ & {$0.43$} & $0.94$ \\
 &{($-0.14$)}&&& { $(0.29)$}  &  &  &  &  &  \\
    \hline
$f_p$ & $0.01$ & $0.006$ & $0.004$ &  $0.043$ &  $0.059$& $0.037$&  $0.62$& {$0.38$}& $0.70$  \\
\hline
\end{tabular}
}
\label{table_perf}
\end{table*}

\begin{figure}[h]
  \centering
  \includegraphics[scale=0.5]{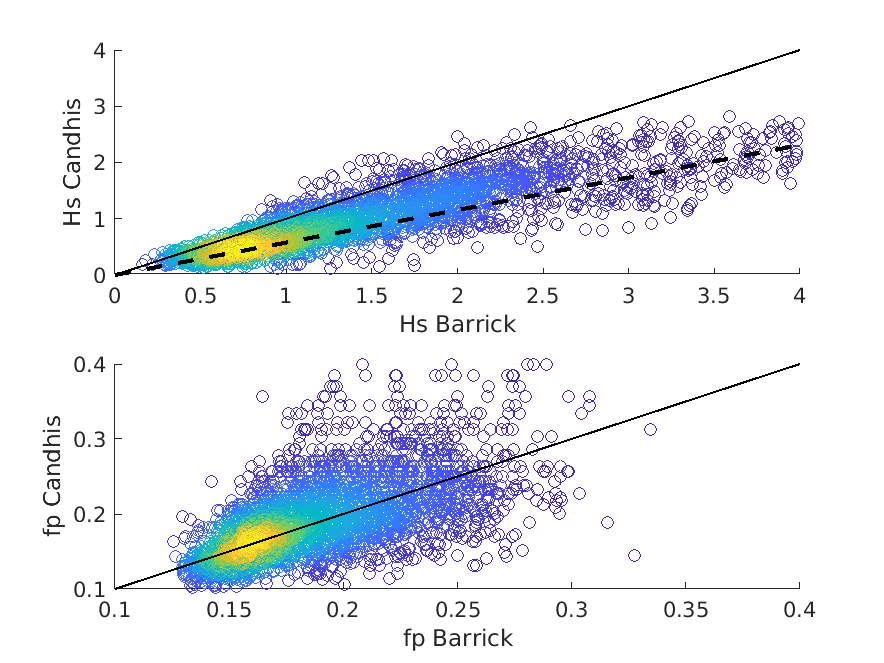}
  \caption{Scatter plot of radar $H_s$ and $f_p$ from Barrick  method versus CANDHIS buoy where each point is colored by the spatial density of nearby points. The thin black line is the regression line with unit slope and the thick dashed line is the regression line with slope $1/0.58$ that shows the need to rescale Barrick's estimate. }
  \label{Figure8_scatterplotBarrick_atCandhis}
\end{figure}

\begin{figure}[h]
  \centering
  \includegraphics[scale=0.5]{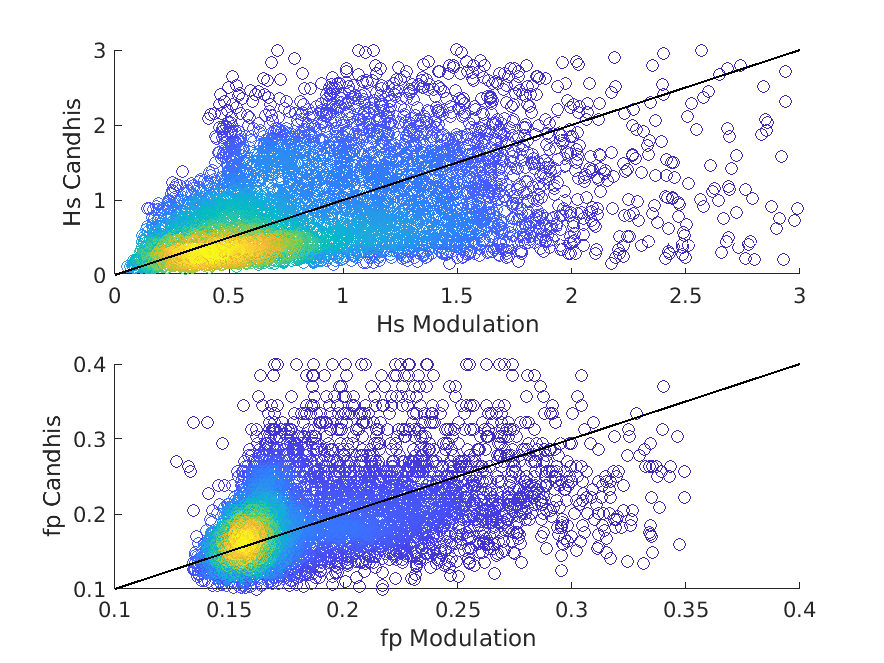}
 \caption{Scatter plot of radar $H_s$ and $f_p$ from Modulation method versus CANDHIS buoy where each point is colored by the spatial density of nearby points. }
  \label{Figure8_scatterplotModulation_atCandhis}
\end{figure}

\subsection{Global comparison with WWIII}
{Although the WWIII model is not a perfect ground truth, it can be used for a coarse evaluation of the radar estimates over the entire radar coverage. For each radar cell and the entire time series, we calculated the RMS difference between the $H_s$ predicted by WWIII and the $H_s$ estimated by either Barrick's or the Modulation method as well as the Pearson correlation coefficient between these quantities. The time coverage was also calculated as the proportion of successful estimates in the time series (after elimination of the outliers). Note that the tests used to select the second-order peaks in Barrick's method results in a lower time coverage for this latter method. 

  Figure \ref{fig:superplotHs} shows the maps obtained for the statistical indices using the two methods, with the calibration factor $0.58$ applied to Barrick's method estimate of $H_s$. The values considered to be irrelevant have been eliminated and shown in white. These are the radar cells for which the time coverage is lower than $30\%$ or the Pearson correlation coefficient is lower than $0.3$. The map confirms that Barrick's method is more accurate than the Modulation method in estimating the $H_s$, with a RMSD of the order of $40$ cm in a large central zone (and probably better if the inaccuracy of WWIII itself is accounted for) out to about $30$ km range. The Pearson correlation coefficient remains larger than about $0.8$ in the same area. The Modulation method has in general a larger RMSD (of the order of $60$ cm at best) and smaller Pearson correlation coefficient (less than $0.7$). \col{Because no constraint was imposed at this stage in the Modulation method (such as e.g. threshold conditions regarding certain quantities), it has by construction a higher spatial and temporal coverage than the Doppler-based method}. The same conclusions apply to the peak wave frequency (figure \ref{fig:superplotfp}) but with smaller values of the Pearson correlation coefficient for the two methods. Note that the modulation method is strongly affected around some ranges (number 15, 25-30 corresponding to distances of about 22 and 37-45 km) by the power grid artifact mentioned in figure \ref{fig:rangedop}. However, relevant estimates are obtained in the far central zone and will now be discussed in more details.}

\begin{figure}[h]
  \centering
	\includegraphics[scale=0.375]{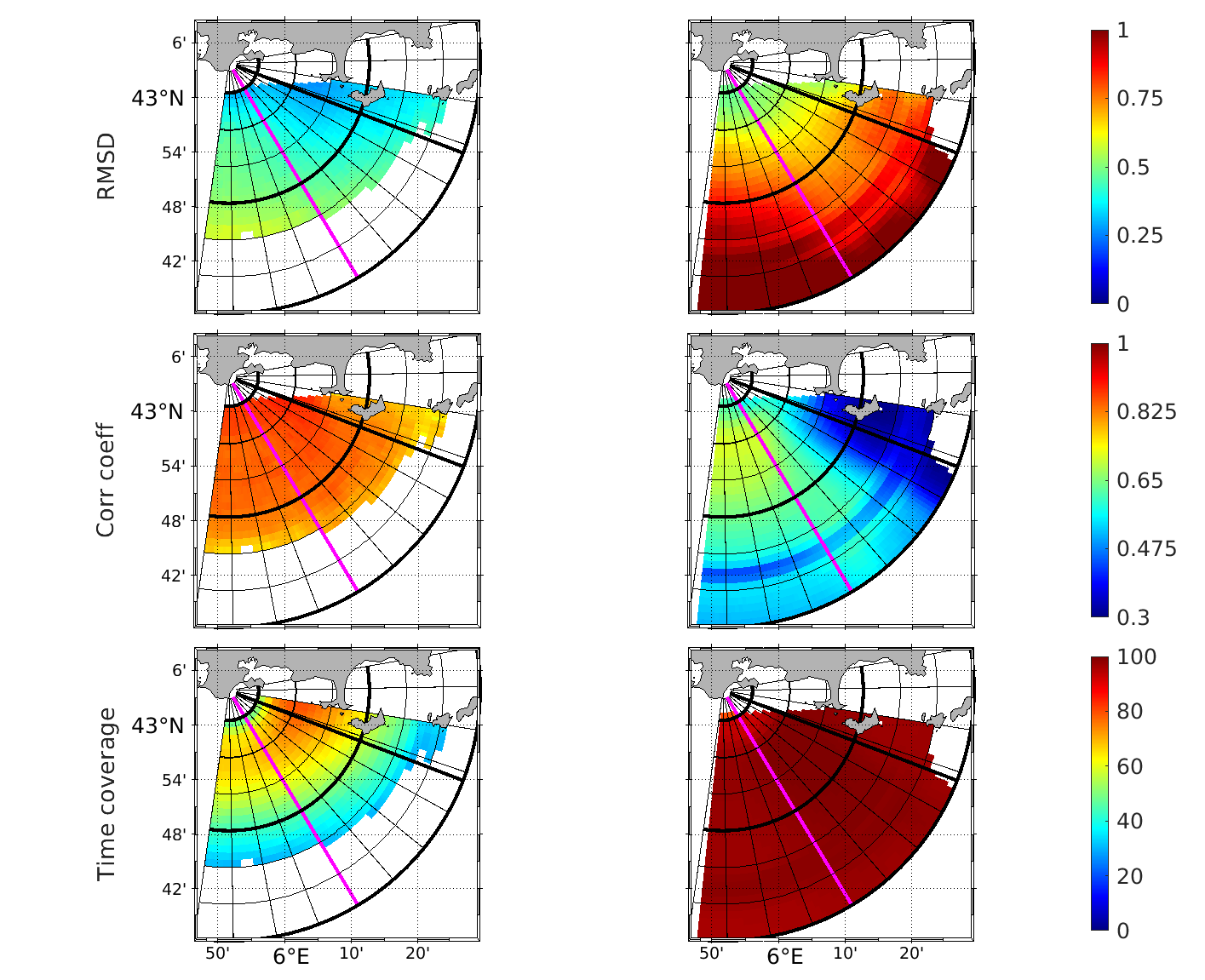}
        \caption{{Spatial maps of the statistical performance indicators for the radar measurement of $H_s$ obtained for the year $2020$ with the calibrated Barrick's method (left column) and the modulation method (second column) when the WWIII is taken as reference: RMSD (in meter)(top); Pearson correlation coefficient (middle); Time coverage (in percent) (bottom). The polar grid has steps of 5 units in range (7.5 km) and 5 degrees in azimuth. The black arcs mark the ranges r5, r20, and r35 ($8$, $30$, and $52.5$ km) respectively; the black radial indicates the azimuth $78$ of the CANDHIS buoy used for the calculations in Table \ref{table_perf} while the magenta indicates azimuth $40$ used for the long range calculation (Section \ref{sec:farrange} below).}}\label{fig:superplotHs}
           \end{figure}

\begin{figure}[h]
  \centering
	\includegraphics[scale=0.375]{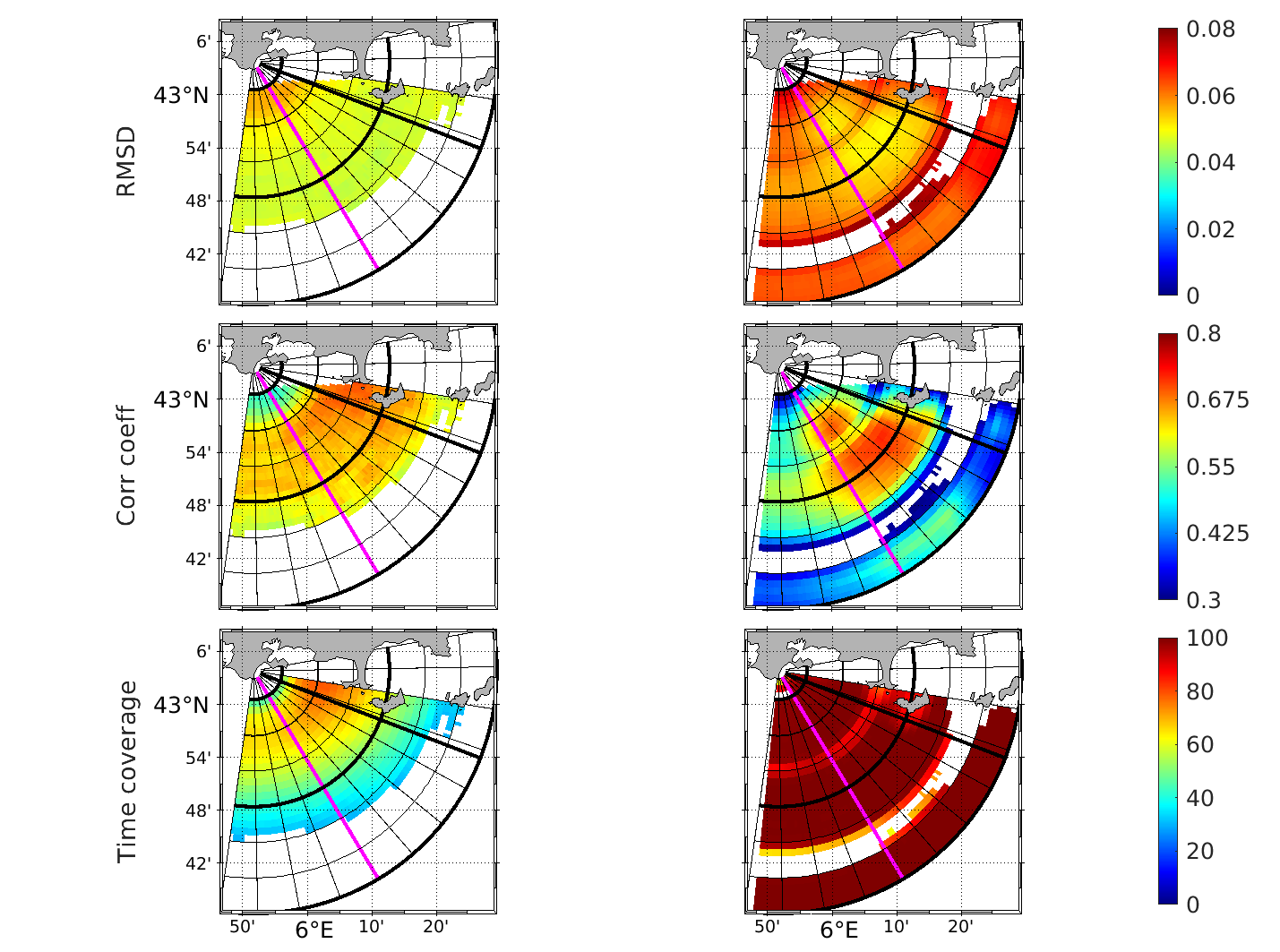}
 \caption{Same as figure \ref{fig:superplotHs} for the peak wave frequency (in Hz).}\label{fig:superplotfp}
 \end{figure}

\subsection{\col{Statistical performances in the long range}}\label{sec:farrange}
\col{We now examine the robustness of the two methods with respect to the range. To do this, we calculated the statistical performance indices along two distinct radials. The first one is the azimuth $78$ of the CANDHIS buoy, which is marked with a thick black line in figure  \ref{fig:superplotHs}; the second is azimuth $40$, which is directed towards the open sea and is marked with a thick magenta line. Figure \ref{fig:perfradial} shows the evolution of the RMSD, Pearson correlation coefficient and time coverage along these two azimuths for the Barrick's and Modulation method, where WWIII is again taken as the reference. As expected, the RMS error for the $H_s$ is increasing with range while the Pearson correlation coefficient is decreasing. However, the RMS errors  obtained with the Modulation method are much larger than those obtained with Barrick's method and the Pearson correlation coefficient is much smaller. Taking as a maximum acceptable error the mean annual value of $H_s$ calculated at each range with the WWIII model, we see that the $H_s$ estimates obtained with the modulation method are relevant up to a distance of about 22 km along the azimuth 78 (CANDHIS direction) and 46 km along the azimuth $40$ (open sea). In spite of this increasing RMS error, the Pearson correlation coefficient obtained with the Modulation method remains meaningful (say, greater than $0.5$) over large distances, especially in the open sea.
Regarding the $f_p$ estimate, the method performances are less contrasted, especially in the open sea. Their RMS errors show little dependence upon range and remain much smaller than the mean parameter value. The Pearson correlation remains larger than about $0.6$ up to a distance of $35$ km.
  To better evaluate the quality of estimation in the long range, we plotted the time series of the estimated $H_s$ and $f_p$ at the cell range $32$ ($48$ km) (figure \ref{fig:timeseries_longrange}). The $H_s$ estimates with the Modulation method follow the overall trend of variation of the WWIII model but suffer from important errors;  the $f_p$ estimates are very weakly correlated to the WWIII data and unable to follow their full variability. On the contrary Barrick's method yields sparse but relevant estimates, especially for the $H_s$. This is confirmed by the scatter plots in figure 
             \ref{fig:scatterplot_longrange}, which shows large scatter of estimates with the Modulation method while the rare estimates obtained with Barrick's method are well aligned with the WWIII values. It must be noted, however, that the selection tests that have been employed to identify the transition between the first- and second-order Doppler spectrum have made this last method more reliable. Again, it must be mentioned that we did not impose any test criterion to the Modulation method while evaluating the wave parameters. This explains at the same time its much higher time coverage and its much higher level of error.}

\begin{figure}[h]
  \centering
  \includegraphics[scale=0.4]{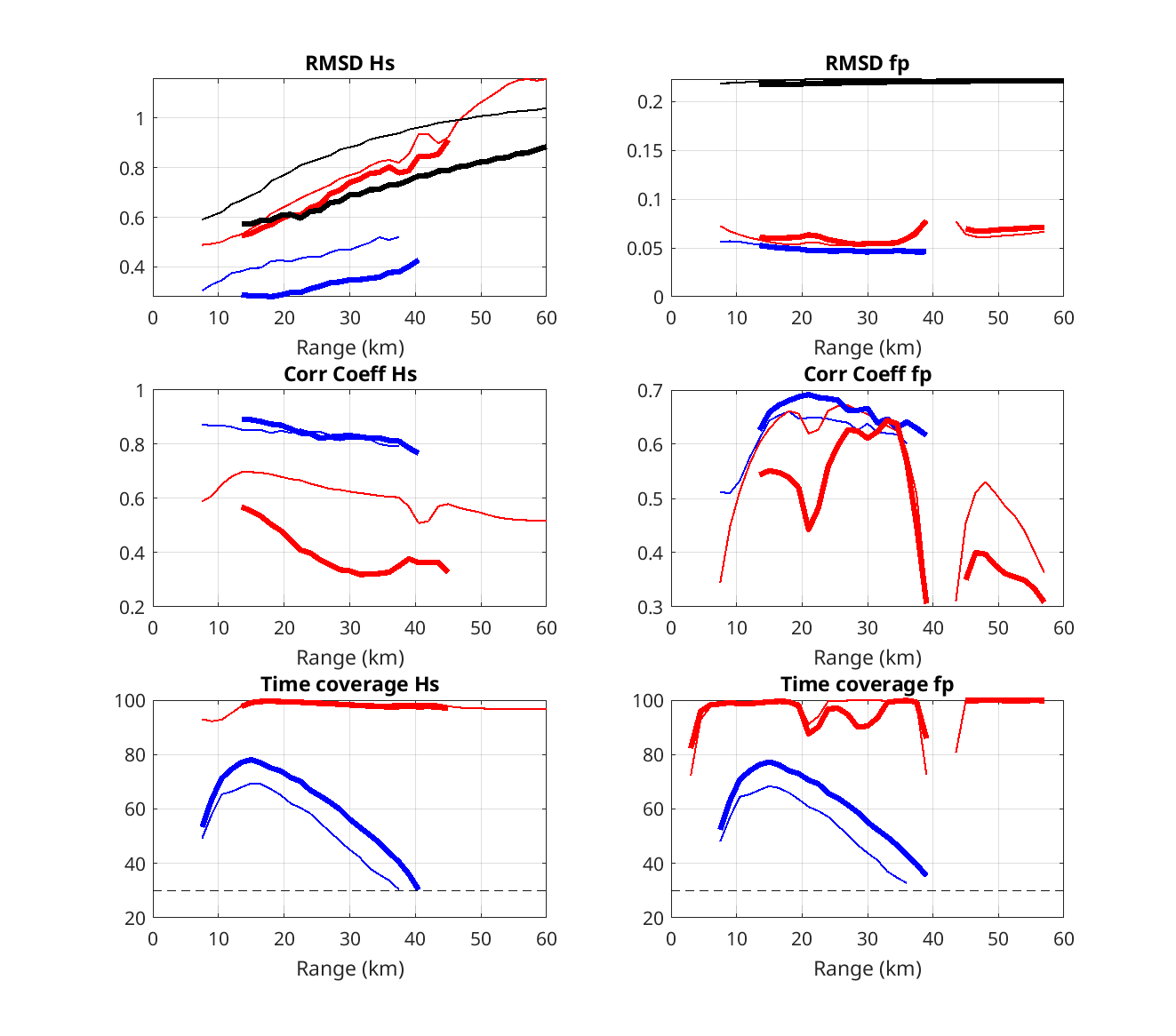}
  \caption{{Statistical performance indicators for Barrick's method (blue lines) and the Modulation method (red lines). The black lines represent the mean values of $H_s$ and $f_p$
      according to the WWIII model at each range cell. The thick lines correspond to the azimuth $78$ (CANDHIS) while the thin line correspond to the azimuth $40$ (open sea).}}
  \label{fig:perfradial}

 \end{figure}

\begin{figure}[h]
  \centering
  \includegraphics[scale=0.14]{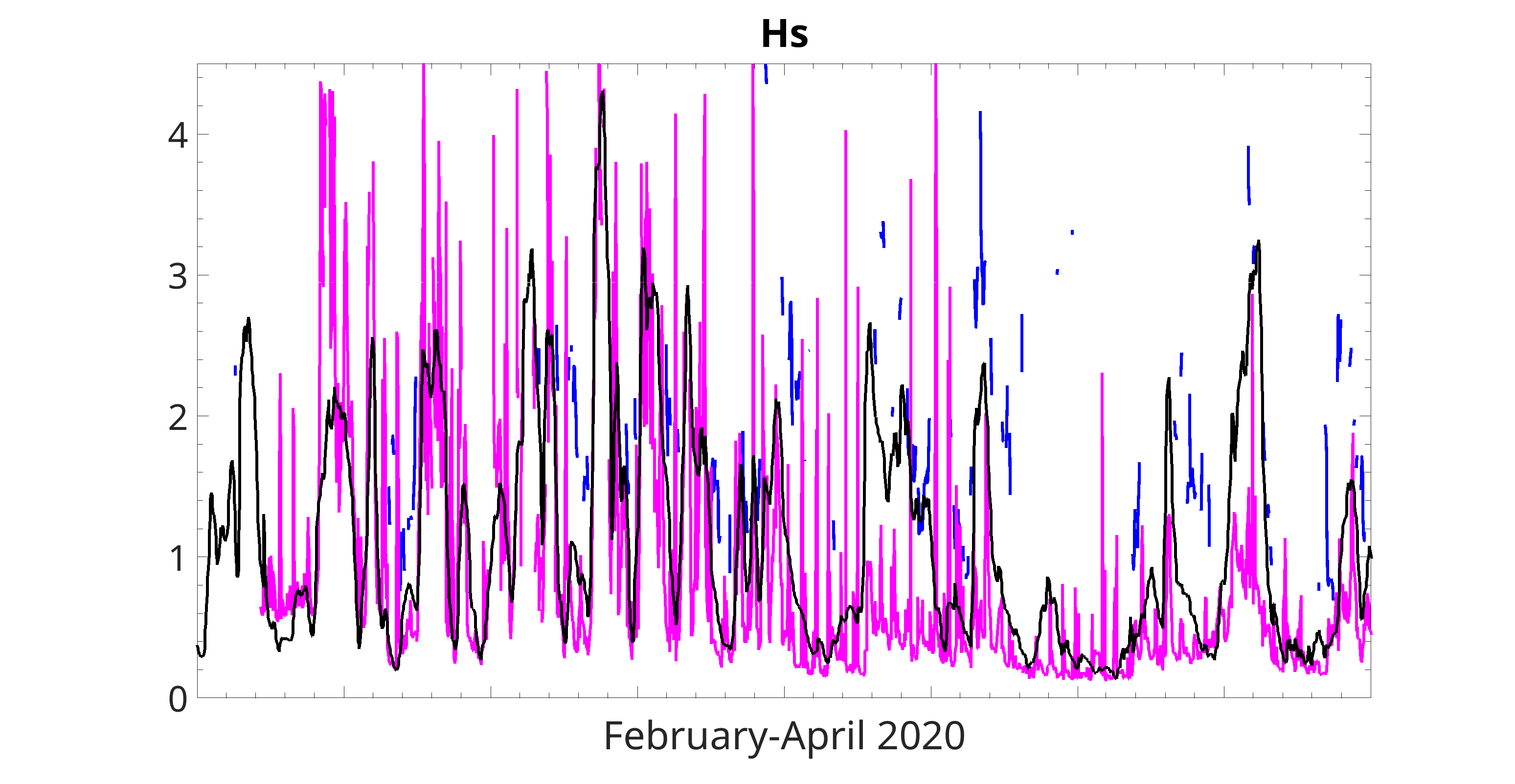} 
  \includegraphics[scale=0.14]{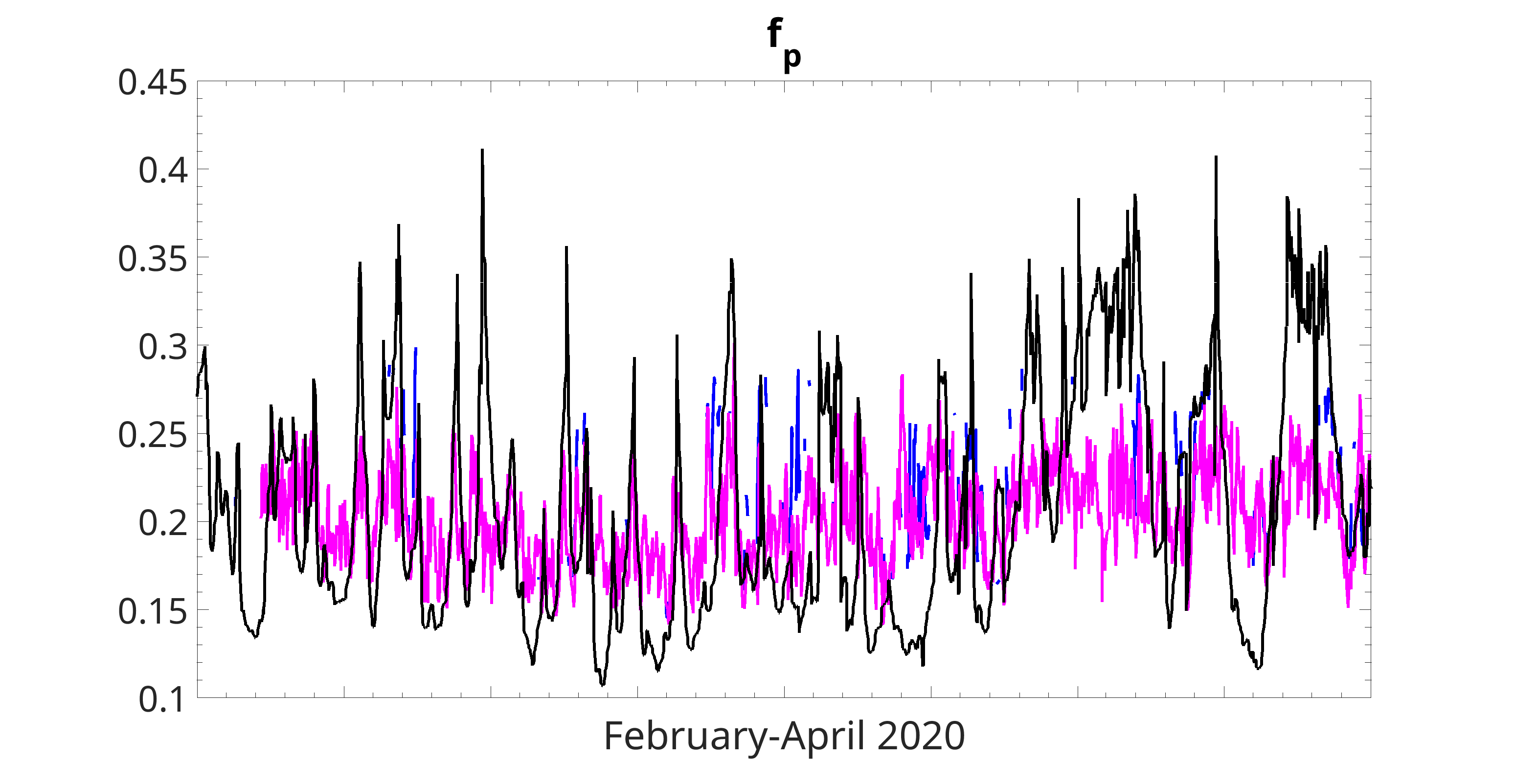}
  \caption{{Time series of $H_s$ (top) and $f_p$ (bottom) in the long range ($48$ km) along the azimuth $40$ (magenta line of figure  \ref{fig:superplotHs}) as estimated by the WWIII model (red lines), Barrick's method (blue symbols) or the Modulation method (magenta symbols).}}
    \label{fig:timeseries_longrange}
\end{figure}

\begin{figure}[h]
  \centering
  \includegraphics[scale=0.5]{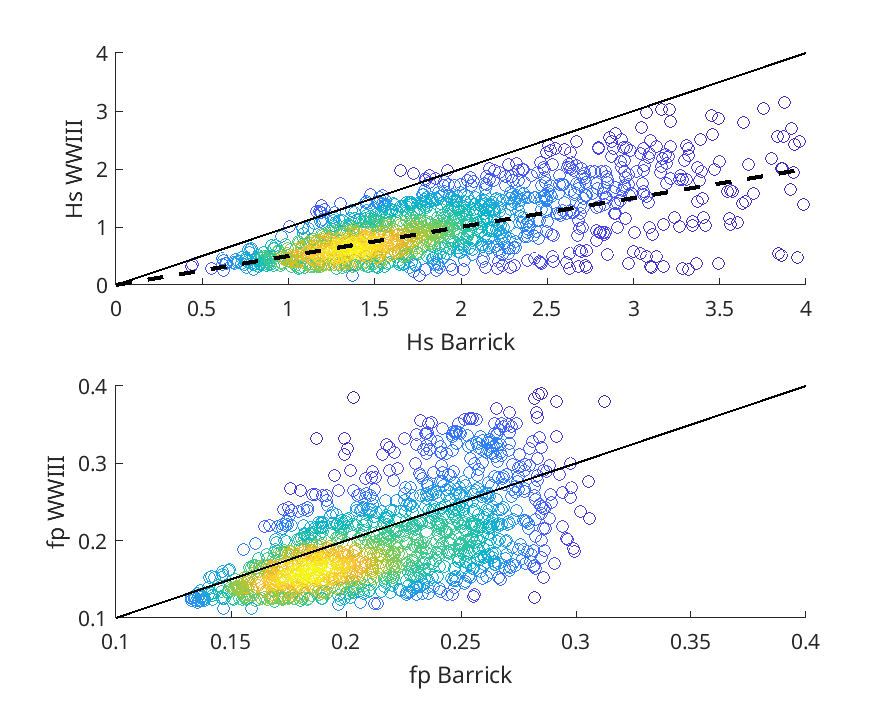}
  \includegraphics[scale=0.5]{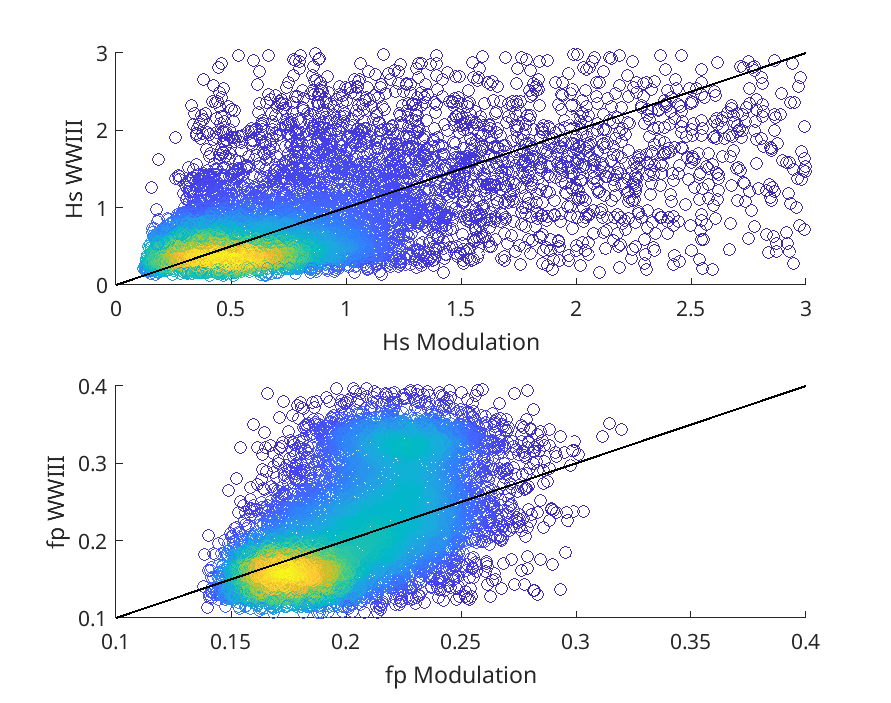}
  \caption{\col{Scatter plots of  $H_s$ and $f_p$ in the long range ($48$ km) along the azimuth $40$ (magenta line of figure  \ref{fig:superplotHs}) as estimated with Barrick's method (top) and the Modulation method (bottom) and compared with the WWIII model.}}
    \label{fig:scatterplot_longrange}
\end{figure}

\subsection{Discussion}
{Barrick's method, which is based on the second-order Doppler spectrum, has well-known limitations, which we summarize below. The roughness parameter $K_0H_s$, where $K_0$ is the electromagnetic wave number, should not be too large for otherwise the perturbation expansions in the scattering orders would break down. This is clearly satisfied in the present case since at $16.15$ MHz with the local wave conditions ($H_s<4$ m), we have $K_0H_s<1.2$
  Another condition for the validity of the asymptotic formulae \ref{eq:Hs} and \ref{eq:fp} (see \cite{barrick_RS77}) is that the peak wave frequency of the ocean wave spectrum be much lower than the Bragg frequency, that is $f_p<<f_B$, a condition that can be violated at low sea states. This explains why the quality of the estimation deteriorates as the mean frequency increases, as has already been mentioned in Section \ref{sec:perfCANDHIS}. In addition, the signal-to-noise ratio of the second-order Doppler spectrum increases with the sea state and may be insufficient in the case of a calm sea. For these reasons, Barrick's method is not adapted for measuring small values of $H_s$ nor large values of $f_p$, a situation which in general corresponds to low wind speed. Beyond these conditions, Barrick's approach is obviously limited in range because it requires the Doppler spectrum side lobes to be larger than the noise level. As a result, the estimation of wave parameters also has a shorter range than the estimation of current, which relies on the sole first-order Bragg peak.
  The limitations of the Modulation method are less obvious to establish. However, by construction this method also requires a clear separation of scales between the Bragg wave and the dominant ocean waves, which again excludes low sea states and peak wave frequencies near the Bragg frequency ($2.5$ seconds). The heuristic justification of the long-wave modulation is limited to small long wave steepness but does not a priori impose any condition on the roughness parameter $K_0H_s$. It is therefore expected to be robust at large wave heights. As it uses the full received voltage and not only its second-order part, the method is also expected to be able to cover a greater range than the Doppler-based method, as long as the slow amplitude modulation is visible. Since the latter is more pronounced when longer waves are more developed and have larger height and periods, the modulation method is expected to be favoured by high winds and to work better in the open sea than in coastal aeras. This is partly confirmed by our experimental study. \col{However, at this early stage, the modulation method is not able to challenge Barrick's method in any respect. Although it produces meaninful estimates at near and middle range and show some correlation with the WWIII model at long range, it is far too inaccurate at the moment to be a useful complement to Barrick's method. In addition, it suffers from some experimental artifacts (which are specific to our installation) such as the power grid instability, which prevent us from using the method over the full range of distances.}
\subsection{Comparison with the literature}

A number of studies have been published in the literature on the estimation of peak wave heights and frequency from single \col{or multiple} HFR stations and a single radar frequency. Some of them reported systematic comparisons with in situ instruments over a significantly long period (at least several months) using either phased array \citep{Wyatt_CoastEng03,ramos_JAOT09, Gomez_Oceans2015,lopez_JTECH16,lopez_JMSE19,alattabi_JTECH19,alattabi_JTECH21} or compact radar systems \citep{roarty_IEEE15,zhou_GRSL15,long_JounSens11,atan_OceanEng16,saviano_Frontiers20,basanez_RemSen20}. All of these methods are based on the exploitation of the second-order Doppler spectrum using either Barrick's original methods, the full wave spectrum inversion method, or empirical methods developed by the radar manufacturers. The performance obtained varies considerably depending on the geographical location, the radar frequency, \col{the number of stations}, the measurement range, the antenna array, the inversion method chosen, the data post-processing choices, etc. However, two general trends can be identified within this diversity of results. First, the estimation of $H_s$ with phased array systems is generally slightly better than that with compact systems, with high Pearson correlation coefficients ($>0.9$) and RMSD often in the order of 30-40 cm and less than about 60 cm in all studies. The observed RMSD with compact systems are generally larger, in the order of 40-80 cm, with a large scatter of values and in some cases very large RMSD ($>1$ m); the associated Pearson correlation coefficients are generally smaller, mostly in the range of 0.5-0.9, except for the study \cite{atan_OceanEng16}, which reports exceptionally good performances of 18-29 cm RMSD and 0.78-0.86 Pearson correlation coefficient. However, the decision to eliminate the $>25$\% outliers (when compared with buoy measurement) before calculating the statistics is questionable. The second point is that the estimation of the peak frequency is less systematic and generally of lower quality than that of $H_s$, with Pearson correlation coefficients of the order of 0.5-0.7.

{Our results are in agreement with these trends. The RMSD and Pearson correlation coefficient obtained for the estimation of $H_s$ with Barrick ranges from $0.3$ to about $0.5$ m and $0.8$ to $0.9$, respectively, for a wide angular sector in the near and medium range (up to about 35 km, see figure \ref{fig:superplotHs}c). The estimation of $f_p$ is much coarser, with a lower Pearson correlation coefficient ($0.6-0.7$ and an RMSD of at best $0.045$ (that is a relative error of the order of $20$\%). However, these performances are based on a comparison with the WWIII which implies an extra error. The actual performances when compared with actual in situ measurements would be probably better.}

\section{Conclusion}\label{sec:discussion}

In this work, we have tested the accuracy of the classical Barrick's method for estimating the main sea state parameters based on weighted integrals of the second-order backscattered Doppler spectrum. For this purpose, we analyzed a one-year HFR data set from a monostatic station in the Toulon region. The results presented in this study have been compared with the data collected by the CANDHIS buoy located near the island of Porquerolles {as well as hourly data simulated from the high-resolution WWIII model.} For the classical Barrick's method, the results obtained for the SWH have a good accuracy compared to the time series of the CANDHIS buoy, with a very high correlation and a low RMSD value between both of them (close to $90\%$ and around $0.3$ m, respectively, at $32$ km from the radar). {A scaling factor of $0.58$ was found to be necessary in order to adjust the statistics of the two data sets and to further reduce the RMSD. With regard to the estimation of the peak wave frequency, this method has some shortcomings in capturing high values, due to insufficient scale separation between the long wave and the Bragg wave.}

{In addition, a new method, called the ``Bragg Wave Modulation Method'' (the ``Modulation Method'' for short), was also introduced and evaluated together with Barrick's method. This technique uses the amplitude modulation of the received voltage time series and does not assume a decomposition of the signal in perturbation orders. It was found to be much less accurate than Barrick's method at this stage \col{and needs to be complemented with further quality tests}. A further limitation of this new method, at least for the $H_s$, is that it requires a preliminary calibration procedure with some ground truth measurements. This is not the case with Barrick's method, which is based on the ratio of the first and second-order Doppler spectra and is therefore insensitive to absolute calibration ({although a rather universal scaling factor improves its accuracy}). The  ``Modulation Method'' will be pursued and hopefully improved in future works. }

\section*{Acknowledgment}Part of this research and the first author (VMM) have been funded by the Agence Nationale de la Recherche (ANR) under grant ANR-22-ASTR-0006-01 (ROSMED project: ``Radar à Ondes de Surface en MEDiterranée'').  We acknowledge the MOOSE program (Mediterranean Ocean Observing System for the Environment) coordinated by CNRS-INSU and the Research Infrastructure ILICO (CNRS-IFREMER). We thank the Parc National de Port-Cros (PNPC), the ``Association Syndicale des Propri\'etaires du Cap B\'enat'' (ASPCB) as well as the Group Military Conservation (GMC) and the Marine Nationale for hosting our radar installation in Porquerolles, Cap Bénat and Fort Peyras, respectively. In addition, we would like to thank Alejandro C\'aceres-Euse for providing the WWIII simulation results in the study region.

\end{document}